\documentclass[preprint,showpacs,preprintnumbers,amsmath,amssymb]{revtex4}

\usepackage{graphicx}
\usepackage{epsfig}
\usepackage{amsmath}
\usepackage{graphics}

\begin{document}

\title{
Reflection and Refraction of Bose-condensate Excitations 
}

\author{Shohei Watabe,$^{1}$ and Yusuke Kato$^{2}$} 
\affiliation{
$^{1}$
Department of Physics, University of Tokyo, Tokyo,  113-0033, Japan}
\affiliation{
$^{2}$
Department of Basic Science, University of Tokyo, Tokyo, 153-8902, Japan}

%\date{}

\begin{abstract}
We investigate the transmission and reflection of Bose-condensate excitations 
in the low energy limit across a potential barrier 
separating two condensates with different densities. 
The Bogoliubov excitation in the low energy limit has 
the incident angle where the perfect transmission occurs. 
This condition corresponds to the Brewster's law for the electromagnetic wave. 
The total internal reflection of the Bogoliubov excitation 
is found to occur at a large incident angle 
in the low energy limit. 
The anomalous tunneling named by Kagan {\it et al.} 
[Yu. Kagan {\it et al.}, Phys. Rev. Lett., {\bf 90}, 130402 (2003)] 
can be understood in terms of the impedance matching. 
%In the case of the normal incidence, 
%our result in the low energy limit is consistent 
%with that for the weakly interacting one-dimensional Bose gases 
%treated as a Tomonaga-Luttinger liquid. 
In the case of the normal incidence, 
comparison with the results in Tomonaga-Luttinger liquids is made. 
\end{abstract}


\maketitle

\section{Introduction}

%The realization of Bose-Einstein condensation in atomic gases~\cite{Anderson1995, Davis1995} 
%has allowed us to investigate properties of Bogoliubov excitations experimentally. 
%Using nondestructive phase-contrast imaging, 
%it was found that the speed of sound induced by modifying the trap potential 
%was consistent with Bogoliubov theory~\cite{Andrews1997}. 
%Using the Bragg scattering, 
%the static structure factor of a condensed Bose gas was measured in the phonon regime~\cite{Stamper-Kurn1999}. 
%It was also reported that the excitation spectrum agreed 
%with the Bogoliubov spectrum using the Bragg scattering~\cite{Steinhauer2002}. 
%At the moment, 
%the transmission and reflection of Bogoliubov excitations 
%are one of the issues that has not been investigated yet experimentally. 

The transmission and the 
reflection are basic concepts in physics, 
that have been studied for strings, elastic bodies, fluids, 
electromagnetic waves, quantum particles, Tomonaga-Luttinger liquids and so on. 
The present paper is devoted 
to the study of the transmission and the reflection of 
collective excitations in the condensed Bose system. 
This proceeds from the earlier study of 
the so-called {\it anomalous tunneling}~\cite{Kagan2003}, 
which is described as follows: 
Bogoliubov excitations transmit perfectly 
through the potential barrier separating two condensates 
in the low energy limit. 
This is in markedly contrast to single particle tunneling 
in the usual quantum mechanics, 
where the perfect reflection occurs in the low energy limit. 

We review a rough sequence of studies of the anomalous tunneling. 
Kovrizhin initially discussed the transmission coefficient 
of Bogoliubov excitations through 
the $\delta$-function potential barrier~\cite{Kovrizhin2001}. 
Kagan {\it et al.} also studied the tunneling of Bogoliubov excitations 
through the rectangular potential~\cite{Kagan2003}. 
%to study the motion of excitations 
%at the background of the inhomogeneous condensate density~\cite{Kagan2003}. 
Through the study in Ref.~\cite{Kagan2003}, 
Kagan {\it et al.} pointed out that 
this phenomenon is anomalous, in comparison to the usual quantum tunneling of a single particle. 

Danshita {\it et al.} considered the transmission of Bogoliubov excitations 
across the $\delta$-function barrier 
separating two condensates with different macroscopic phases~\cite{Danshita2006}. 
They found that the anomalous tunneling disappears 
when the phase difference reaches the critical value 
giving the maximum supercurrent of the condensate. 
%Through this work, 
%they concluded that one of the origins of the anomalous tunneling 
%is the localized components near the potential barrier 
%with imaginary momenta. 

%Turning our eyes to the weakly interacting one-dimensional Bose gases, 
%the excitation spectrum is consistent with the Bogoliubov excitation~\cite{Lieb}. 
%In the long wavelength limit, this system can be described by the Luttinger liquids. 
%When one considers the multiple reflections, 
%the perfect transmission occurs in the symmetrical system. 
%We shall compare this results with the result in this paper in the case of the normal incident.   

These works were done using the one-dimensional potential. 
On the other hand, 
the problem of Bogoliubov excitations scattered by a spherical potential 
was examined~\cite{Padmore1972,FujitaMThesis,FujitaUnpublished}. 
The energy dependence of the cross section 
of the low energy scattering by the spherical potential 
is consistent with the Rayleigh scattering of a sound wave 
in the classical wave mechanics, 
{\it i.e.,} the cross section vanishes 
in the low energy limit~\cite{FujitaMThesis,FujitaUnpublished}. 

Using the fact that 
the wave function of the excited state in the low energy limit corresponds 
to the macroscopic wave function of the condensate~\cite{Fetter1972}, 
%Answering the question whether the anomalous tunneling occurs or not 
%with the potential barrier having the general form, 
Kato {\it et al.} concluded that the anomalous tunneling occurs 
for a potential barrier $V(x)$ being arbitrary symmetric function ($V(-x) = V(x)$)~\cite{Kato2007}. 
They also indicated that the perfect tunneling appears even at finite temperatures 
within the scheme of the Popov approximation~\cite{Kato2007}. 

%Recently, Tsuchiya and Ohashi 
%studied tunneling properties of Bogoliubov phonons 
%%in condensed Bose system 
%taking notice of the quasi-particle current 
%near the potential barrier~\cite{Tsuchiya2008}. 
%They found that the quasi-particle current increases near the potential barrier 
%inducing the supercurrent counterflow. 
%They proposed an interpretation that this increase explains the increase of the transmission coefficient of the Bogoliubov phonon. 
%In order to conserve the total current, 
%they use the Gross-Pitaevskii equation added in the anomalous average. 
%However, this formulation brings the gapful excitation, 
%and they did not confirm whether the anomalous tunneling occurs or not within their formulation. 
%It is considered that the gapless excitation is necessary to cause the anomalous tunneling, 
%so that their discussion is in a dilemma. 
%The problem, 
%whether the anomalous tunneling occurs or not 
%in the formulation satisfying the number conservation and the gapless excitation, 
%still remains. 

In the Tomonaga-Luttinger liquids, on the other hand, 
the perfect transmission has been also discussed in the context of quantum wire, 
using the renormalization-group~\cite{Kane1992}, 
and considering the multiple reflections~\cite{Safi1995}. 
Especially, using the renormalization-group theory, 
it was shown that 
the barrier is an irrelevant perturbation in an attractive fermion system~\cite{Kane1992}. 
That is to say, 
the perfect transmission is known to occur in Tomonaga-Luttinger liquids in superconducting phase. 

%
%The present paper is also devoted 
%to compare the perfect transmission of the Bogoliubov phonon 
%with that of Tomonaga-Luttinger liquids. 

%Returning to the subject, 
%we note that all earlier studies have considered tunneling problems of Bogoliubov excitations 
%across the symmetric potential barrier separating two condensates. 
%They have ever considered only symmetrical systems, 
%and hence we have ever seen one of the aspects of the tunneling phenomenon of collective excitations. 
%All essential physics and implications 
%about the anomalous tunneling were not also exposed to us. 
%When we consider more general problem, {\it i.e.}, 
%the tunneling problem of Bogoliubov phonon through the potential barrier 
%separating two condensates with different densities, 
%the phenomenon of the anomalous tunneling would occur as the limiting case of the {\it symmetrical} system. 
%Further, from the general problem, 
%we can have more implications 
%about the tunneling problem of Bogoliubov phonon. 
%In these circumstances, 
%we investigate the tunneling problem of Bogoliubov phonon through the arbitrary potential barrier 
%separating two condensates with different densities. 
%We shall also consider that 
%the incident Bogoliubov phonon does not only 
%run toward the wall perpendicularly 
%but also runs toward the wall with an arbitrary incident angle. 
%This extended problem is analogous to the problem of 
%the reflection and refraction of the wave 
%at the interface between two different mediums. 

As seen in the above, 
the property of the transmission of the Bogoliubov excitation in the low energy limit 
has not only a specific character of the condensed Bose system 
but also something in common with that of the classical wave mechanics 
and that of the Tomonaga-Luttinger liquids. 
From this point of view, 
we have three aims in this paper as follows: 
(i) exposing novel phenomena of the transmission and the reflection of the Bogoliubov phonon, 
(ii) giving a physical implication about the anomalous tunneling 
analogous to the classical wave mechanics,   
(iii) comparing the result of the tunneling problem of the Bogoliubov phonon with 
that of the Tomonaga-Luttinger liquids. 

To achieve these aims, 
we investigate the tunneling problem of the Bogoliubov phonon through the arbitrary potential barrier 
separating two condensates with different densities. 
We consider that 
the incident Bogoliubov phonon does not only 
run toward the wall perpendicularly 
but also runs toward the wall with an arbitrary incident angle. 
This extended problem is analogous to the problem of 
the reflection and the refraction of a wave 
at the interface between two different mediums.

We may summarize the results of this paper as follows: 
(i) The Bogoliubov phonon has an incident angle 
where the perfect transmission occurs 
between condensates with different densities. 
This condition corresponds to the Brewster's law, {\it i.e.}, 
the sum of the incident angle and the refracted angle is equal to $\pi/2$. 
There also exists the total internal reflection of the Bogoliubov excitation in the low energy limit. 
(ii) The perfect transmission of the Bogoliubov excitation 
in the low energy limit 
can be regarded as a result of the impedance matching 
between equivalent condensates separated by the potential wall. 
The impedance is inversely proportional to the sound speed of the Bogoliubov phonon. 
(iii) At normal incidence, 
our result in the low energy limit is consistent 
with the result obtained by the theory of Tomonaga-Luttinger liquids~\cite{Safi1995}, 
when one uses the Luttinger liquid parameter 
for the weakly interacting Bose gases. 
We note that 
the negative density reflection in the interacting condensed Bose system cannot 
be necessarily identified with the Andreev reflection. 

The outline of this paper is as follows. 
In Section II, we shall give a formulation of the problem, 
and derive the transmission and reflection coefficients. 
To obtain these coefficients, 
we use properties that the wave function of the excited state in the low energy limit 
corresponds to the condensate wave function, and the constancy of the energy flux. 
In Section III, we shall discuss results in accordance with three aims. 
We also propose future problems on the experimental and theoretical sides. 
In Section IV, we summarize our results. 

\section{Formulation and Results}
In this section, 
we derive the transmission and reflection coefficients of the Bogoliubov excitation 
in the low energy limit. 
We use the mean-field theory, say, the Gross-Pitaevskii equation~\cite{Gross1961,Pitaevskii1961} 
and the Bogoliubov equation~\cite{Bogoliubov1947}. 
To discuss the property of the Bogoliubov excitation in the low energy limit, 
we use the fact that 
the wave function of the excited state in the low energy limit corresponds 
to the macroscopic wave function of the condensate~\cite{Fetter1972}. 
To evaluate the transmission and reflection coefficients, 
we use the constancy of the energy flux. 

\begin{figure}
\begin{center}
\includegraphics[width=8cm,height=8cm,keepaspectratio,clip]{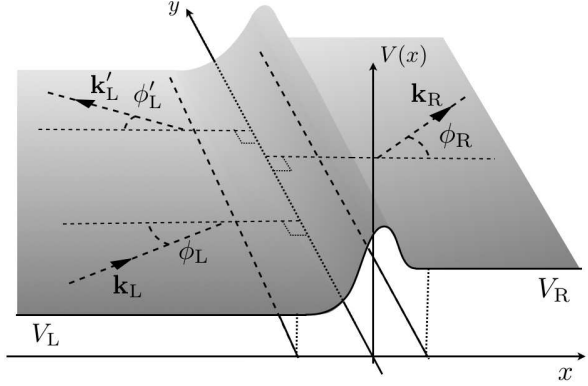}
\end{center}
\caption{
A schematic figure of the system. 
In this figure, we take $xy$ plane as an incident plane. 
}
\label{potential.fig}
\end{figure}

The stationary Gross -Pitaevskii equation written in dimensionless form 
is given by 
\begin{eqnarray}
\hat{\mathcal H}\overline{\Psi}(\overline{{\bf r}}) = 0, 
\label{GPeq}
\end{eqnarray}
where 
\begin{eqnarray}
\hat{\mathcal H} \equiv -\frac{1}{2}\overline{\nabla}^{2}
+\overline{V}(\overline{\bf r})-1+|\overline{\Psi}(\overline{{\bf r}})|^{2}, 
\label{Hamil}
\end{eqnarray}
and the Bogoliubov equation also written in dimensionless form is given by 
\begin{eqnarray}
\left(
\begin{array}{cc}
\hat{h}' & -{\overline{\Psi}(\overline{{\bf r}})}^2 \\
{\overline{\Psi}^{*}(\overline{\bf r})}^{2} & -\hat{h}'
\end{array}
\right) 
\left(
\begin{array}{cc}
{u}(\overline{\bf r}, \overline{\varepsilon}) \\ {v}(\overline{\bf r}, \overline{\varepsilon}) 
\end{array}
\right)
= \overline{\varepsilon}
\left(
\begin{array}{cc}
{u}(\overline{\bf r};\overline{\varepsilon}) \\
{v}(\overline{\bf r};\overline{\varepsilon})
\end{array}
\right), 
\label{Bogo}
\end{eqnarray}
where $\hat{h}' \equiv \hat{\mathcal H} + |\overline{\Psi}(\overline{{\bf r}})|^{2} $. 
We have introduced the following notations 
$\overline{\bf r} \equiv {\bf r}/\xi$, 
$\overline{\nabla} \equiv \xi\nabla$
$\overline{\Psi}(\overline{{\bf r}}) \equiv \sqrt{g/\mu}\Psi({\bf r})$, 
$\overline{V}(\overline{\bf r}) \equiv V({\bf r})/\mu$, 
and $\overline{\varepsilon} \equiv \varepsilon/\mu\varepsilon$, 
where $g (> 0)$ is the coupling constant of the two-body short-range interaction, 
$\mu$ is the chemical potential, 
and $\xi$ is the healing length defined by $\xi \equiv \hbar/\sqrt{m\mu}$ 
with $m$ being the mass. 
Henceforth, 
we omit the bar for simplicity. 
We take the potential $V({\bf r})$ to be a function of $x$ 
and allow $V(x)$ to be asymmetric $V(x) \neq V(-x)$. 
We consider $V(x)$ being superposition of 
short-range potential near $x = 0$ 
and a potential step with asymptotic form: 
\begin{eqnarray}
V(x) \rightarrow \left \{ 
\begin{array}{ll}
V_{\rm L}, &\qquad {\rm for} \quad x \ll -1
\\
V_{\rm R}, &\qquad {\rm for} \quad x\gg 1, 
\end{array}
\right .
\label{Pot}
\end{eqnarray} 
where both $V_{\rm L}$ and $V_{\rm R}$ are smaller than unity. 
The profile of $V(x)$ is shown in Fig.~\ref{potential.fig}. 

\begin{figure}
\begin{center}
\includegraphics[width=8cm,height=9cm,keepaspectratio,clip]{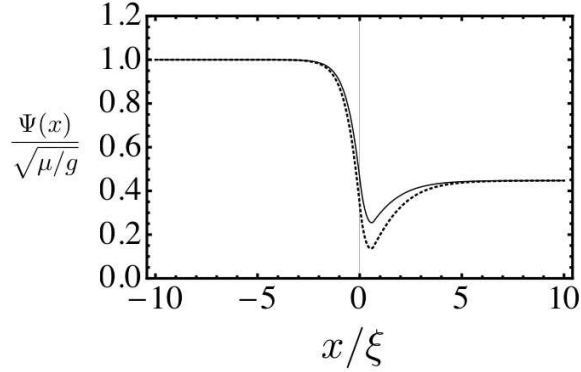}
\end{center}
\caption{ 
The numerical solutions of the condensate wave function. 
We use the potential energy as the function given in Eq. (\ref{PotNum}). 
A solid line represents the case for $V_{\rm b} = 3$, and a dotted line represents the case for $V_{\rm b} = 5$. 
}
\label{Fig2.fig}
\end{figure}

There are steady-states of a Bose-Einstein condensate 
%in terms of the mean-field 
even in the presence of a potential step~\cite{Seaman2005}. 
%According to Ref.~\cite{Seaman2005}, 
%there are solutions with the {\it trivial phase} which is spatially constant, 
%and with the {\it nontrivial phase} which varies spatially accompanied with supercurrent. 
%we shall consider a solution giving the minimum energy 
%among the steady-states, and hence 
In this paper, 
we shall treat a case without a supercurrent 
where the phase is spatially constant. 
Henceforth, $\Psi({\bf r})$ is considered to be real. 
In this situation, 
we have the following asymptotic behavior 
of the condensate wave function: 
\begin{eqnarray}
\Psi(x) \rightarrow \left \{ 
\begin{array}{ll}
\sqrt{1-{V}_{\rm L}}, &\qquad {\rm for} \quad x \ll -1
\\
\sqrt{1-{V}_{\rm R}}, &\qquad {\rm for} \quad x\gg 1. 
\end{array}
\right .
\label{CondWaveAsym}
\end{eqnarray} 
We note that 
our objective is to derive the transmission and reflection coefficients 
which can be adopted to the general case. 
However, we shall give an example here, 
in order to compare our general result with a concrete example. 
For instance, we assume that the potential $V(x)$ 
%in the dimensional form 
has the following behavior: 
\begin{align}
V(x) = 
\left \{ 
\begin{array}{ll}
0, &\qquad {\rm for} \quad x \leq 0, 
\\
V_{\rm b}, &\qquad {\rm for} \quad 0 < x \leq 1/\sqrt{2}, 
\\
0.8, &\qquad {\rm for} \quad 1/\sqrt{2} < x. 
\end{array}
\right .
\label{PotNum}
\end{align} 
We use the strength of the potential barrier $V_{\rm b}$ as $V_{\rm b} = 3$ and $5$. 
In Fig.~\ref{Fig2.fig}, 
we show condensate wave functions obtained by numerical calculations, 
in the presence of the above potential $V(x)$. 
%in the case of the {\it trivial phase}. 
%In Fig.~\ref{Fig2.fig}, 
A solid line represents the numerical result for $V_{\rm b} = 3$, 
and a dotted line represents the numerical result for $V_{\rm b} = 5$. 
These numerical solutions satisfy the asymptotic behavior given in Eq.~(\ref{CondWaveAsym}). 
When the potential energy in the asymptotic regime is less than the chemical potential, 
{\it i.e.,} $V_{\nu} < \mu$ with $\nu = {\rm L}$ for $x \ll -1$ and $\nu = {\rm R}$ for $x \gg 1$, 
there are condensate wave functions spatially constant in the asymptotic regime, 
as shown in Fig.~\ref{Fig2.fig}. 
Hence, Bogoliubov excitations even in the low energy limit exist in both sides of the potential barrier. 
%, and it is allowed to discuss the transmission of the Bogoliubov excitation. 

We shall consider the situation 
where a Bogoliubov excitation with a wavenumber vector ${\bf k}_{\rm L}$ 
runs against the potential wall at an angle $\phi_{\rm L}$ 
with respect to the normal vector of the wall ({\it i.e.}, $x$-direction). 
The Bogoliubov phonon 
is split into a transmitted wave 
with a refraction having a wavenumber vector ${\bf k}_{\rm R}$ at an angle $\phi_{\rm R}$ 
and a reflected wave 
with ${\bf k}_{\rm L}'$ at an angle $\phi_{\rm L}'$ 
as shown in Fig.~\ref{potential.fig}. 
%The aim of this paper is to determine the transmission coefficient $T$ 
%and the reflection coefficient $R$ of the Bogoliubov excitation in the low energy limit. 

Owing to the translational invariance in $y$, $z$ directions, 
the Gross-Pitaevskii equation reduces to 
\begin{eqnarray}
\hat{h}\Psi(x) = 0, 
\label{GPX}
\end{eqnarray} 
with 
\begin{eqnarray}
\hat{h} = 
-\frac{1}{2}\frac{d^{2}}{dx^{2}} 
+{V}(x)-1+\Psi^{2}(x). 
\label{HamilX}
\end{eqnarray} 
The solution of the Bogoliubov equation has the form of 
\begin{eqnarray}
\left(
\begin{array}{cc}
{u}({\bf r};{\varepsilon}) \\
{v}({\bf r};{\varepsilon})
\end{array}
\right)
= 
\exp{(ik_{y}y+ik_{z}z)}
\left(
\begin{array}{cc}
{u}(x;{\varepsilon}) \\
{v}(x;{\varepsilon})
\end{array}
\right), 
\label{solBogoYZ}
\end{eqnarray} 
and the Bogoliubov equation is given by 
\begin{eqnarray}
\left(
\begin{array}{cc}
\hat{h}'' & - \Psi^{2}(x) \\
\Psi^{2}(x) & -\hat{h}''
\end{array}
\right)
\left(
\begin{array}{cc}
{u}(x;{\varepsilon}) \\
{v}(x;{\varepsilon})
\end{array}
\right)
= 
\varepsilon
\left(
\begin{array}{cc}
{u}(x;{\varepsilon}) \\
{v}(x;{\varepsilon})
\end{array}
\right), 
\label{BogoX}
\end{eqnarray}
where 
%\begin{eqnarray}
$
\hat{h}'' = \hat{h} + \Psi^{2} (x) + k_{\perp}^{2}/2
$, 
%\label{Hamil2Dash}
%\end{eqnarray}
and $k_{\perp} = \sqrt{k_{y}^{2} + k_{z}^{2}}$. 

Let us consider asymptotic forms 
of a Bogoliubov mode. 
At $|x|\gg 1$ where $\Psi(x) \approx \sqrt{1-V_{\nu}}$ 
with $\nu = {\rm L\,\, and\,\, R}$, 
the basis of the solution is given by two plane-wave solutions 
\begin{eqnarray}
\left(
\begin{array}{cc}
{u}(x;{\varepsilon}) \\
{v}(x;{\varepsilon})
\end{array}
\right)
= 
\exp{(\pm i k_{||, \nu}x)}
\left(
\begin{array}{cc}
\tilde{u}_{\nu}(\varepsilon) \\
\tilde{v}_{\nu}(\varepsilon)
\end{array}
\right), 
\label{BogoPlane}
\end{eqnarray}
and 
exponentially growing or converging solutions 
\begin{eqnarray}
\left(
\begin{array}{cc}
{u}(x;{\varepsilon}) \\
{v}(x;{\varepsilon})
\end{array}
\right)
= 
\exp{(\pm \kappa_{\nu} x)}
\left(
\begin{array}{cc}
\tilde{v}_{\nu}(\varepsilon)\\
-\tilde{u}_{\nu}(\varepsilon)

\end{array}
\right), 
\label{BogoExp}
\end{eqnarray}
where 
\begin{eqnarray}
\tilde{u}_{\nu}(\varepsilon) 
\equiv 
\sqrt{\frac{\sqrt{(1-V_{\nu})^{2}+\varepsilon^{2}}}{2\varepsilon}+\frac{1}{2}}, 
\qquad\quad
\tilde{v}_{\nu}(\varepsilon)
\equiv
\sqrt{\frac{\sqrt{(1-V_{\nu})^{2}+\varepsilon^{2}}}{2\varepsilon}-\frac{1}{2}}, 
\label{BogoBase}
\end{eqnarray}
satisfying the normalization 
%\begin{eqnarray}
$
|\tilde{u}_{\nu}(\varepsilon)|^{2}-|\tilde{v}_{\nu}(\varepsilon)|^{2} = 1
$.
%\label{BogoNorm}
%\end{eqnarray}
The wavenumber $k_{||, \nu}$ 
and 
the growing rate (or converging rate) $\kappa_{\nu}$ 
are given by, respectively, 
\begin{eqnarray}
\left \{
\begin{array}{lll}
k_{||, \nu} &=& \sqrt{k_{\nu}^{2} - k_{\perp}^{2}} 
\\
\kappa_{\nu} &=& \sqrt{2\left[1-V_{\nu}+\sqrt{\varepsilon^{2}+(1-V_{\nu})^{2}}\right]
+k_{\perp}^{2}}\,\, ,
\end{array}
\right. 
\label{WaveNumber}
\end{eqnarray}
where $k_{\nu} = \sqrt{2\left [ \sqrt{(1-V_{\nu})^{2}+\varepsilon^{2}}-(1-V_{\nu}) \right ]}$.

Wavenumbers $k_{||, \nu}$ and $k_{\perp}$ 
which are, respectively, 
parallel and perpendicular components against the normal vector of the wall 
can be written as 
$k_{||, \nu} = k_{\nu} \cos{\phi_{\nu}}$ and $k_{\perp} = k_{\nu}\sin{\phi_{\nu}}$. 
Owing to the translational invariance in $y$, $z$ directions, 
we have the law of reflection 
\begin{eqnarray}
\phi_{\rm L} = \phi_{\rm L}'. 
\label{lawReflection}
\end{eqnarray} 
Because of the law of reflection, 
there does not exist retroreflection in the weakly interacting condensed Bose system. 
That is to say, 
when the Bogoliubov excitation runs toward the wall 
with an incident angle, the reflected excitation 
does not go back the way one has come. 
%With respect to the incident wave and the transmitted wave, 
We also have a relation between the incident angle and the transmitted angle given by 
\begin{eqnarray}
k_{{\rm L}}\sin{\phi_{{\rm L}}} = k_{{\rm R}}\sin{\phi_{{\rm R}}}. 
\label{SnellLaw1}
\end{eqnarray} 
In the low energy limit, the energy has the relation 
$\varepsilon \approx c_{\nu}k_{\nu}$, 
where $c_{\nu}$ is a sound speed of Bogoliubov phonon in the dimensionless form 
given by $c_{\nu}\equiv \sqrt{1-V_{\nu}}$. 
From the relation in Eq. (\ref{SnellLaw1}), in the low energy limit, 
we recover the Snell's law 
\begin{eqnarray}
\frac{\sin{\phi_{\rm L}}}{c_{\rm L}}
= 
\frac{\sin{\phi_{\rm R}}}{c_{\rm R}}. 
\label{SnellLaw2}
\end{eqnarray}

We assume that an incident wave comes from the left side of 
the potential barrier as shown in Fig.~\ref{potential.fig}. 
Far from the barrier, 
exponentially-diverging components 
$e^{-\kappa_{\rm L} x}$ and $e^{\kappa_{\rm R} x}$ should be absent 
in the physical solution,  
and exponentially-converging components 
$e^{\kappa_{\rm L} x}$ and $e^{-\kappa_{\rm R} x}$
are negligible, 
so that we have the following asymptotic form: 
\begin{eqnarray}
\left(
\begin{array}{cc}
{u}(x;{\varepsilon}) \\
{v}(x;{\varepsilon})
\end{array}
\right)
=
\left \{
\begin{array}{ll}
\left [
a(\varepsilon)\exp{(ik_{||,{\rm L}}x)} + b(\varepsilon)\exp{(-ik_{||, {\rm L}}x)}
\right ]
\left(
\begin{array}{cc}
\tilde{u}_{{\rm L}}(\varepsilon)
\\
\tilde{v}_{{\rm L}}(\varepsilon)
\end{array}
\right),
\qquad 
&
{\rm for} \quad 
x\ll -1
\\
\\
c(\varepsilon)
\exp{(ik_{||, {\rm R}}x)}
\left(
\begin{array}{cc}
\tilde{u}_{{\rm R}}(\varepsilon)
\\
\tilde{v}_{{\rm R}}(\varepsilon)
\end{array}
\right),
\qquad 
&
{\rm for} \quad 
x
\gg 1.
\end{array}
\right.
\label{BogoTR}
\end{eqnarray}

We introduce functions $S(x, \varepsilon)$ and 
$G(x, \varepsilon)$~\cite{Kato2007, FujitaMThesis, Kagan2003, Kovrizhin2001} as 
\begin{eqnarray}
\left \{
\begin{array}{ll}
S(x; \varepsilon) \equiv u(x;\varepsilon) + v(x; \varepsilon)
\\
G(x; \varepsilon) \equiv u(x;\varepsilon) - v(x; \varepsilon).   
\end{array}
\right.
\label{SG}
\end{eqnarray}
According to Eq. (\ref{BogoTR}), 
we have the asymptotic form of the function $S(x;\varepsilon)$ 
in the low energy limit up to order ${\mathcal O}(\varepsilon^{1/2})$
as 
\begin{eqnarray}
\left \{
\begin{array}{llllll}
S(x; \varepsilon) &\sim& 
\varepsilon^{-1/2}
\sqrt{
2(1-V_{\rm L})
}
\left \{ 
a(\varepsilon)+b(\varepsilon)
+ 
i\varepsilon
\gamma_{\rm L}
%\frac{\cos{\phi_{\rm L}}}{\sqrt{1-V_{\rm L}}}
\left [
a(\varepsilon)-b(\varepsilon)
\right ]
x
\right \},
&\qquad& {\rm for} \quad x \ll -1
\\
S(x; \varepsilon) &\sim & 
\varepsilon^{-1/2}
\sqrt{
2(1-V_{\rm R})
}
c(\varepsilon)
\left ( 
1+i\varepsilon
\gamma_{\rm R}
%\frac{\cos{\phi_{\rm R}}}{\sqrt{1-V_{\rm R}}}
x
\right ),
&\qquad& {\rm for} \quad x \gg 1, 
\end{array}
\right.
\label{SAsymp}
\end{eqnarray} 
where we define $\gamma_{\nu} \equiv \cos{\phi_{\nu}}/\sqrt{1-V_{\nu}}$.

From the Bogoliubov equation in Eq. (\ref{BogoX}), 
functions $S(x;\varepsilon)$ and $G(x;\varepsilon)$ satisfy following equations: 
\begin{eqnarray}
\left \{
\begin{array}{ccc}
\displaystyle{ 
\left ( 
\hat{h} + k_{\perp}^{2}/2
\right ) 
}
S(x, \varepsilon) 
&=& \varepsilon G(x, \varepsilon)
\\
\displaystyle{ 
\left ( 
\hat{h}^{(-)} + k_{\perp}^{2}/2
\right ) 
}
G(x, \varepsilon) 
&=& \varepsilon S(x, \varepsilon),
\end{array}
\right.
\label{SGEq}
\end{eqnarray}
where $\hat{h}^{(-)} = \hat{h} + 2 \Psi^{2}(x)$. 
We expand functions $S(x;\varepsilon)$ and $G(x;\varepsilon)$ with the energy $\varepsilon$ as 
\begin{eqnarray}
\left \{
\begin{array}{ccc}
S(x;\varepsilon) &=& 
\varepsilon^{-1/2}
\sum\limits_{n=0}^{\infty} \varepsilon^{n} S^{(n)}(x)
\\
G(x;\varepsilon) &=& 
\varepsilon^{-1/2}
\sum\limits_{n=0}^{\infty} \varepsilon^{n} G^{(n)}(x). 
\end{array}
\right. 
\label{SGExpand}
\end{eqnarray}
We note that 
the factor $\varepsilon^{-1/2}$ 
comes from the normalization in Eq. (\ref{BogoBase}). 
Equations for $S^{(0)}(x)$, $G^{(0)}(x)$, and $S^{(1)}(x)$ are 
given by 
\begin{eqnarray}
\left \{
\begin{array}{ccc}
\hat{h}S^{(0)}(x) &=& 0
\\
\hat{h}^{(-)}G^{(0)}(x) &=& 0
\\
\hat{h}S^{(1)}(x) &=& G^{(0)}(x). 
\end{array}
\right.
\label{S0G0S1}
\end{eqnarray}

First, we consider the solution $G^{(0)}$. 
The operator $\hat{h}^{(-)}$ has a second-order ordinary differential. 
The solution of $G^{(0)}$ can be written by 
\begin{eqnarray}
G^{(0)}(x) &=& B_{\rm I} g_{\rm I}(x) + B_{\rm II} g_{\rm II}(x), 
\label{G0}
\end{eqnarray} 
with constants $B_{\rm I}$, $B_{\rm II}$ and 
two independent solutions $g_{\rm I}(x)$ and $g_{\rm II}(x)$ satisfying, respectively,  
\begin{eqnarray}
\left \{
\begin{array}{ccc}
\hat{h}^{(-)}g_{\rm I}(x) = 0, 
\quad g_{\rm I}(x=0) = 1, 
\quad 
\displaystyle{ 
\left. \frac{d g_{\rm I}(x)}{dx} \right |_{x=0} = 0
}
\\
\\
\hat{h}^{(-)}g_{\rm II}(x) = 0, 
\quad g_{\rm II}(x=0) = 0, 
\quad 
\displaystyle{ 
\left. \frac{d g_{\rm II}(x)}{dx} \right |_{x=0} = 1. 
}
\end{array}
\right. 
\label{BaseG}
\end{eqnarray} 
At $|x| \gg 1$, 
the asymptotic behavior of $\hat{h}^{(-)}$ can be written as 
$-\frac{d^{2}}{2dx^{2}} + 2(1-V_{\nu})$. 
Hence, 
any linear combination of $g_{\rm I}(x)$ and $g_{\rm II}(x)$ generally diverges 
exponentially at $|x| \gg 1$. 
On the other hand, 
$G^{(0)}$ should not have any exponentially diverging terms 
at $|x| \gg 1$. 
Thus, it is allowed that we set $B_{\rm I} = B_{\rm II} = 0$ 
as in Ref.~\cite{Kato2007}, 
and hence we do not have to treat the special solution of 
the inhomogeneous equation of the third equation 
in Eq. (\ref{S0G0S1}).

We shall consider general solutions of homogeneous equations given by 
\begin{eqnarray}
\hat{h}S^{(0)}(x) = 0, \qquad \hat{h}S^{(1)}(x) = 0. 
\label{EqS1}
\end{eqnarray} 
Equations in Eq.~(\ref{EqS1}) are second-order linear differential equations, 
and hence 
solutions $S^{(0)}(x)$ and $S^{(1)}(x)$ 
can be written 
as 
\begin{eqnarray}
\left \{
\begin{array}{ccc}
S^{(0)} = 
A_{\rm I}^{(0)} s_{\rm I}(x) + A_{\rm II}^{(0)} s_{\rm II}(x)
\\
S^{(1)} = A_{\rm I}^{(1)} s_{\rm I}(x) + A_{\rm II}^{(1)} s_{\rm II}(x), 
\end{array}
\right. 
\label{SoluS0S1}
\end{eqnarray}
by using two independent solutions $s_{\rm I}(x)$ and $s_{\rm II}(x)$ satisfying 
\begin{eqnarray}
\left \{
\begin{array}{ccc}
\displaystyle{
\hat{h}s_{\rm I}(x) = 0, 
\quad s_{\rm I} (x=0)= 1, 
\quad \left. \frac{d s_{\rm I}(x)}{dx} \right |_{x=0} = 0
}
\\
\\
\displaystyle{
\hat{h}s_{\rm II}(x) = 0, 
\quad s_{\rm II}(x=0) = 0, 
\quad \left. \frac{d s_{\rm II}(x)}{dx} \right |_{x=0} = 1. 
}
\end{array}
\right. 
\label{BaseSISII}
\end{eqnarray}

At $|x|\gg1$, the asymptotic behavior of $\hat{h}$ is described 
as $\hat{h} \sim -\frac{1}{2}\frac{d^{2}}{dx^{2}}$, 
so that we 
have solutions as 
\begin{eqnarray}
\left \{
\begin{array}{ccc}
s_{\rm I}(x) &\sim& \alpha_{{\rm I}, \nu} + \beta_{{\rm I}, \nu}x 
\\
s_{\rm II}(x) &\sim& \alpha_{{\rm II}, \nu} + \beta_{{\rm II}, \nu}x,  
\end{array}
\right.
\label{sIMAsym}
\end{eqnarray}
with $\nu = {\rm L}$ for $x \ll -1$ 
and $\nu = {\rm R}$ for $x \gg 1$.

As a result, 
we have the asymptotic behavior of the solution $S(x;\varepsilon)$ 
given by 
\begin{eqnarray}
S(x;\varepsilon)& \sim& 
\varepsilon^{-1/2}
\left [
A_{\rm I}(\varepsilon)\alpha_{{\rm I}, \nu}
+A_{\rm II}(\varepsilon)\alpha_{{\rm II}, \nu}
\right ]
\nonumber
\\
&&+\varepsilon^{-1/2}[
A_{\rm I}(\varepsilon)\beta_{{\rm I}, \nu}
+A_{\rm II}(\varepsilon)\beta_{{\rm II}, \nu}
]x+{\mathcal O}(\varepsilon^{3/2} ),
\label{AsymS}
\end{eqnarray} 
with $\nu = {\rm L}$ and ${\rm R}$, 
where we define 
$A_{\rm I} (\varepsilon) \equiv 
A_{\rm I}^{(0)} + \varepsilon A_{\rm I}^{(1)}$ and 
$A_{\rm II} (\varepsilon) \equiv 
A_{\rm II}^{(0)} + \varepsilon A_{\rm II}^{(1)}$ 
and 
use Eq.~(\ref{sIMAsym}). 

Let us use the property 
that 
the wave function of the excited state in the low energy limit corresponds 
to the macroscopic wave function of the condensate~\cite{Fetter1972}. 
%Using the solutions $s_{\rm I}(x)$ and $s_{\rm II}(x)$, 
%we express the solution of the Gross-Pitaevskii equation. 
Considering solutions $s_{\rm I,II}(x)$ satisfying 
$\hat{h}s_{\rm I,II}(x) = 0$, 
the solution of the Gross-Pitaevskii equation $\hat{h}\Psi (x) = 0$ 
can be written as 
\begin{eqnarray}
\Psi(x) = \Psi_{0} s_{\rm I}(x) + \Psi_{0}' s_{\rm II} (x), 
\label{SoluGPSISII}
\end{eqnarray}
where $\Psi_{0} \equiv \Psi (x=0)$ 
and $\Psi_{0}' \equiv \left. \frac{d\Psi (x)}{dx} \right |_{x = 0}$. 
Using asymptotic forms for $s_{\rm I}(x)$ and $s_{\rm II}(x)$ in Eq. (\ref{sIMAsym}), 
we evaluate asymptotic behaviors of the condensate wave function as 
\begin{eqnarray}
\Psi(x) \sim (\Psi_{0}\alpha_{{\rm I}, \nu} + \Psi_{0}'\alpha_{{\rm II}, \nu})
+ 
(\Psi_{0}\beta_{{\rm I}, \nu} + \Psi_{0}'\beta_{{\rm II}, \nu})x, 
\label{AsymCondM}
\end{eqnarray}
with $\nu = {\rm L}$ and ${\rm R}$.

We know that asymptotic behaviors of the condensate wave function 
are given in Eq. (\ref{CondWaveAsym}), 
so that 
we have constraints for coefficients $\alpha_{\rm I, \nu}$, $\alpha_{\rm II, \nu}$, 
$\beta_{\rm I, \nu}$, and $\beta_{\rm II, \nu}$ given by 
\begin{eqnarray}
\left \{
\begin{array}{lll}
\Psi_{0}\alpha_{{\rm I}, \nu} + \Psi_{0}'\alpha_{{\rm II}, \nu} &=& \sqrt{1-V_{\nu}}
\\
\Psi_{0}\beta_{{\rm I}, \nu} + \Psi_{0}'\beta_{{\rm II}, \nu} &=& 0, 
\end{array}
\right.
\label{ConstAlphaBeta}
\end{eqnarray}
where $\nu = {\rm L}$ and ${\rm  R}$. 
The second equation of Eq.~(\ref{ConstAlphaBeta}) 
shows that 
vectors $^{\rm t}(\beta_{{\rm I}, \nu}, \beta_{{\rm II}, \nu})$ and 
$^{\rm t}(\Psi_{0}, \Psi_{0}')$ are orthogonal. 
Thus, we write the vector $^{\rm t}(\beta_{{\rm I}, \nu}, \beta_{{\rm II}, \nu})$ 
as 
\begin{eqnarray}
\left (
\begin{array}{lll}
\beta_{{\rm I}, \nu}
\\
\beta_{{\rm II}, \nu}
\end{array}
\right )
\equiv
\tilde{\beta}_{\nu}
\left (
\begin{array}{c}
\Psi_{0}'
\\
-\Psi_{0}
\end{array}
\right ), 
\label{BetaRep}
\end{eqnarray} 
with a constant $\tilde{\beta}_{\nu}$.

Comparing Eq. (\ref{SAsymp}) and Eq. (\ref{AsymS}), 
we have equations for coefficients ${A}_{\rm I, II}(\varepsilon)$, 
the amplitude transmission coefficient $t(\varepsilon) \equiv c(\varepsilon)/a(\varepsilon)$, 
and the amplitude reflection coefficient $r(\varepsilon) \equiv b(\varepsilon)/a(\varepsilon)$, 
\begin{eqnarray}
\left \{
\begin{array}{llllll}
{A}_{\rm I}(\varepsilon)\alpha_{\rm I, L} 
+ 
{A}_{\rm II}(\varepsilon)\alpha_{\rm II, L} 
&= &
\sqrt{2(1-V_{\rm L})} 
a(\varepsilon)
[
1+r(\varepsilon)
]
+ {\mathcal O}(\varepsilon^{2})
\\
{A}_{\rm I}(\varepsilon)\beta_{\rm I, L} 
+ 
{A}_{\rm II}(\varepsilon)\beta_{\rm II, L} 
&= &
i\sqrt{2}
a(\varepsilon)
[
1-r(\varepsilon)
]
\varepsilon\cos{\phi_{\rm L}} 
+ {\mathcal O}(\varepsilon^{2}),
\end{array}
\right.
\label{ATRM}
\end{eqnarray} 
and 
\begin{eqnarray}
\left \{
\begin{array}{llllll}
{A}_{\rm I}(\varepsilon)\alpha_{\rm I, R} 
+ 
{A}_{\rm II}(\varepsilon)\alpha_{\rm II, R} 
&= &
\sqrt{2(1-V_{\rm R})}
a(\varepsilon)
t(\varepsilon)
+ {\mathcal O}(\varepsilon^{2})
\\
{A}_{\rm I}(\varepsilon)\beta_{\rm I, R} 
+ 
{A}_{\rm II}(\varepsilon)\beta_{\rm II, R} 
&= &
i\sqrt{2}
a(\varepsilon)
t(\varepsilon)
\varepsilon\cos{\phi_{\rm R}}
+ {\mathcal O}(\varepsilon^{2}). 
\end{array}
\right. 
\label{ATRP}
\end{eqnarray}

Using Eqs. (\ref{ConstAlphaBeta}), (\ref{BetaRep}), (\ref{ATRM}), and (\ref{ATRP}), 
we have the amplitude transmission coefficient $t(\varepsilon)$ 
and the amplitude reflection coefficient $r(\varepsilon)$ in the low energy limit $\varepsilon \rightarrow 0$ 
\begin{eqnarray}
\left \{
\begin{array}{lll}
\lim\limits_{\varepsilon\rightarrow 0}t(\varepsilon) &= &
\displaystyle{
\frac{2\zeta_{\rm R}}{\zeta_{\rm R}+ \zeta_{\rm L}}
}
\\
\\
\lim\limits_{\varepsilon\rightarrow 0}
r(\varepsilon) &= &
\displaystyle{
\frac{\zeta_{\rm R} - \zeta_{\rm L}}
{\zeta_{\rm R} + \zeta_{\rm L}}
}, 
\end{array}
\right.
\label{lowETR}
\end{eqnarray}
where we define $\zeta_{\nu}\equiv \tilde{\beta}_{\nu}/\cos{\phi_{\nu}}$.

We shall consider the relation between 
$\zeta_{\rm L}$ and $\zeta_{\rm R}$ 
on the basis of constancy of energy flux. 
According to Ref.~\cite{Kagan2003}, 
one has the energy flux averaged over the time in dimensionless form 
given by 
\begin{eqnarray}
\langle 
{\bf Q} ({\bf r} ) 
\rangle
&=& 
\frac{\varepsilon}{2}
{\rm Im}
\left [ 
S^{*}({\bf r}) \nabla S({\bf r}) 
+
G^{*}({\bf r})\nabla G({\bf r})
\right ]. 
\label{EFlux}
\end{eqnarray} 
We shall consider the asymptotic behavior of the energy flux parallel to the normal vector of the wall 
by using Eq. (\ref{BogoTR}). 
%defining $\langle Q_{||,{\rm L}} \rangle \equiv \langle Q_{||} (x \ll -1) \rangle$ 
%and 
%$\langle Q_{||,{\rm R}} \rangle \equiv \langle Q_{||} (x \gg 1) \rangle$. 
The asymptotic behavior of the energy flux in the $x$-direction for $x \ll -1$ is 
given by 
\begin{eqnarray}
\langle Q_{||,{\rm L}} \rangle 
\sim 
{\varepsilon}
\left ( 
|\tilde{u}_{+, {\rm L}}|^{2}+|\tilde{v}_{+, {\rm L}}|^{2} 
\right )
k_{||, {\rm L}}
\left [ 1-|r(\varepsilon)|^{2} \right ]
|a(\varepsilon)|^{2}. 
\label{QL}
\end{eqnarray}
On the other hand,  
we have the asymptotic form of the energy flux 
in the $x$-direction for $x \gg 1$ given by 
\begin{eqnarray}
\langle Q_{||, {\rm R}} \rangle 
\sim 
{\varepsilon}
\left ( 
|\tilde{u}_{+, {\rm R}}|^{2}+|\tilde{v}_{+, {\rm R}}|^{2} 
\right )
k_{||, {\rm R}}
|t(\varepsilon)|^{2}
|a(\varepsilon)|^{2}. 
\label{QR}
\end{eqnarray}
In the low energy limit, 
the energy flux can be written as 
\begin{eqnarray}
\left \{
\begin{array}{lll}
\langle Q_{||, {\rm L}} \rangle 
&\sim  & 
\varepsilon 
c_{\rm L}
\cos{\phi_{\rm L}}
\left [ 
1 - 
|r(\varepsilon)|^{2} 
\right ] 
|a(\varepsilon)|^{2} 
\\
\langle Q_{||, {\rm R}} \rangle 
&\sim  &
\varepsilon 
c_{\rm R}
\cos{\phi_{\rm R}}
|t(\varepsilon)|^{2} 
|a(\varepsilon)|^{2}. 
\end{array}
\right.
\label{lowEQLR}
\end{eqnarray}

Regarding the constancy of the energy flux 
$\langle Q_{||, {\rm L}} \rangle =\langle Q_{||, {\rm R}} \rangle$ 
as the constraint for the variables $\zeta_{\rm L}$ and $\zeta_{\rm R}$, 
we obtain the ratio 
%the relation in the low energy limit given by 
%\begin{eqnarray}
%Z_{\rm L}
%\cos{\phi_{\rm L}}
%c_{\rm L}
%\cos{\phi_{\rm L}}
%\left [
%1-|r(\varepsilon\rightarrow 0)|^{2}
%\right ] 
%= 
%\frac{
%c_{\rm R}
%\cos{\phi_{\rm R}}
%}
%{
%c_{\rm L}
%\cos{\phi_{\rm L}}
%}
%|t(\varepsilon\rightarrow 0)|^{2}.  
%\label{ImpedanceConserve}
%\end{eqnarray} 
%Equation (\ref{ImpedanceConserve}) can be regarded as the constraint 
%for the variables $\zeta_{\rm L}$ and $\zeta_{\rm R}$. 
%Imposing the constraint in Eq. (\ref{ImpedanceConserve}) on Eq. (\ref{lowETR}), 
%we have the ratio 
\begin{eqnarray}
\frac{\zeta_{\rm L}}{\zeta_{\rm R}} = 
\frac{c_{\rm R}}{c_{\rm L}}
\frac{\cos{\phi_{\rm R}}}{\cos{\phi_{\rm L}}}. 
\label{BLBR}
\end{eqnarray}
As a result, 
we have the amplitude transmission coefficient $t(\varepsilon)$ 
and the amplitude reflection coefficient $r(\varepsilon)$ 
in the low energy limit 
as 
\begin{eqnarray}
\lim\limits_{\varepsilon\rightarrow 0}
t(\varepsilon) = 
\frac{2
c_{\rm L}\cos{\phi_{\rm L}}
}{
c_{\rm L}\cos{\phi_{\rm L}}
+
c_{\rm R}\cos{\phi_{\rm R}}
}
, 
\qquad \qquad 
\lim\limits_{\varepsilon\rightarrow 0}
r(\varepsilon) =  
\frac{
c_{\rm L}\cos{\phi_{\rm L}}
-
c_{\rm R}\cos{\phi_{\rm R}}
}
{
c_{\rm L}\cos{\phi_{\rm L}}
+
c_{\rm R}\cos{\phi_{\rm R}}
}.
\label{TRZ}
\end{eqnarray}
These amplitude transmission coefficient $t$ 
and amplitude reflection coefficient $r$ 
satisfy the relation: $1+r=t$, 
as seen in the classical wave mechanics. 
The transmission coefficient $T(\varepsilon)$ 
and the reflection coefficient $R(\varepsilon)$ 
are, respectively, given by ratios 
of the transmitted energy flux and of the reflected energy flux to 
the incident energy flux. 
In the low energy limit, 
$T(\varepsilon)$ and $R(\varepsilon)$ 
are, respectively, given by 
\begin{eqnarray}
\begin{array}{llllll}
\lim\limits_{\varepsilon\rightarrow 0}
T(\varepsilon) &= &
%\begin{displaystyle}
%\frac{
%c_{\rm R}
%\cos{\phi_{\rm R}}
%}
%{
%c_{\rm L}
%\cos{\phi_{\rm L}}
%}
%%|t(\varepsilon\rightarrow 0)|^{2}
%\end{displaystyle}
%&= &
\begin{displaystyle}
\frac{4
c_{\rm L}c_{\rm R}
\cos{\phi_{\rm L}}\cos{\phi_{\rm R}}
}{
(
c_{\rm L}\cos{\phi_{\rm L}}
+
c_{\rm R}\cos{\phi_{\rm R}}
)^{2} 
}
\end{displaystyle}
\\
\\ 
\lim\limits_{\varepsilon\rightarrow 0}
R(\varepsilon) 
&= &
%\begin{displaystyle}
%\frac{
%c_{\rm L}
%\cos{\phi_{\rm L}}
%}
%{
%c_{\rm L}
%\cos{\phi_{\rm L}}
%}
%|r(\varepsilon\rightarrow 0)|^{2}
%\end{displaystyle}
%&=  &
\begin{displaystyle}
\frac{
(
c_{\rm L}\cos{\phi_{\rm L}}
-
c_{\rm R}\cos{\phi_{\rm R}}
)^{2}
}
{
(
c_{\rm L}\cos{\phi_{\rm L}}
+
c_{\rm R}\cos{\phi_{\rm R}}
)^{2}
}, 
\end{displaystyle}
\end{array}
\label{TRZ2}
\end{eqnarray}
satisfying the relation: 
$T(\varepsilon)+R(\varepsilon)=1$. 

Equation (\ref{TRZ2}) and the Snell's law in Eq. (\ref{SnellLaw2}) 
describe the transmission and the reflection of 
the Bogoliubov excitation in the low energy limit. 
We note that $T(\varepsilon\rightarrow 0)$ and $R(\varepsilon\rightarrow 0)$ in Eq. (\ref{TRZ2}) 
do not depend on the detail of $V(x)$ near $x=0$; they depend on $V(x)$ 
only through the asymptotic values $V_{\rm L}$ and $V_{\rm R}$. 
The perfect transmission $T(\varepsilon \rightarrow 0)=1$ 
and $R(\varepsilon\rightarrow 0)=0$ follows from $c_{\rm L} = c_{\rm R}$, 
or equivalently $V_{\rm L}=V_{\rm R}$. 
This result is consistent with earlier studies \cite{Kagan2003,Danshita2006,Kato2007,Kovrizhin2001}.  
The transmission and reflection coefficients in Eq. (\ref{TRZ2}) are 
similar to but different from those of the classical sound mechanics, 
%However, we note that this expression is different from ones, 
as discussed later.

\begin{figure}
\begin{center}
\includegraphics[width=8cm,height=8cm,keepaspectratio,clip]{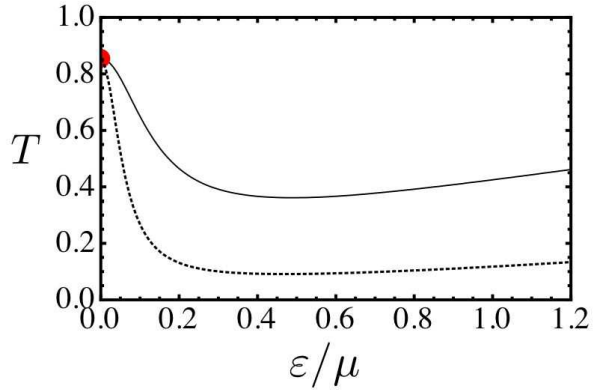}
\end{center}
\caption{
(Color online) 
The transmission coefficient $T$ as a function of the energy $\varepsilon$,  
where the incident angle $\phi_{\rm L} = 0$.  
In the numerical calculation, 
we assume the potential energy given in Eq. (\ref{PotNum}). 
The solid and dotted lines in Fig.~\ref{TE08.fig} are the numerical result in the case for $V_{\rm b} = 3$ and $5$.  
The dot is our analytical prediction in the low energy limit. 
}
\label{TE08.fig}
\end{figure}

The formulation and the result up to the present 
is adopted to the existence of the potential $V(x)$ having the general form. 
Here, let us compare the transmission coefficient of our result with that obtained by the numerical calculation 
in the presence of a potential $V(x)$. 
We shall adopt the potential energy $V(x)$ given in Eq. (\ref{PotNum}). 
In Fig.~\ref{TE08.fig}, 
we show transmission coefficients $T$ in the low energy limit given in Eq. (\ref{TRZ2}) 
and obtained by numerical calculations as a function of the energy $\varepsilon$. 
We set the incident angle as $\phi_{\rm L} = 0$. 
%and use the condensate wave function given in Fig.~\ref{Fig2.fig}. 
Solid and dotted lines in Fig.~\ref{TE08.fig} are numerical results 
in the case for $V_{\rm b} = 3$ and $5$, respectively. 
The dot is our analytical result in the low energy limit. 
We note that 
the transmission coefficient obtained by numerical calculations 
reaches our analytical prediction as the incident energy goes to zero as seen in Fig.~\ref{TE08.fig}. 
This result means that the transmission and reflection coefficients in the low energy limit are 
independent of the potential barrier. These coefficients depend only on asymptotic values of the potential energy.

%In this section, we discuss the physical implications of Eq.~(\ref{TRZ2}). 
%First 
%Note that we do not assume $V(x)=V(-x)$ in the present paper, in contrast to Ref.~\cite{Kato2007}. 

\section{Discussion} 
In this section, 
we shall discuss the property of transmission and reflection of the Bogoliubov phonon. 
First, we shall show the phenomena 
%of the transmission and reflection 
of the Bogoliubov phonon 
having something in common with the electromagnetic wave. 
These are existences of the Brewster's angle and the total internal reflection. 
Second, we shall give an interpretation of the anomalous tunneling in terms of the impedance matching. 
Third, we show the relation between the weakly interacting three-dimensional Bose gas and the one-dimensional Bose gas 
treated as the Tomonaga-Luttinger liquids. 
We also discuss the Andreev-{\it like} reflection, critically. 
Finally, we shall propose future problems on the experimental and theoretical sides. 

\subsection{Brewster's Angle and Total Internal Reflection}

In addition to the ``ordinary'' anomalous tunneling discussed 
in Refs.~\cite{Kagan2003,Danshita2006,Kato2007,Kovrizhin2001}, 
we find another condition 
where we do have the perfect transmission of 
the Bogoliubov excitation in the low energy limit. 
When the incident Bogoliubov phonon has the incident angle $\phi_{\rm L,B}$ 
defined by 
\begin{eqnarray}
\tan{\phi_{\rm L,B}} \equiv {c_{\rm L}}/{c_{\rm R}}, 
\label{Brewster}
\end{eqnarray}
we obtain the perfect transmission: $T = 1$ and $R = 0$. 
This incident angle $\phi_{\rm L,B}$ corresponds to the Brewster's angle 
for the electromagnetic wave, 
satisfying 
the relation $\phi_{\rm L,B} + \phi_{\rm R} = \pi/2$~\cite{LandauElectromagnetic}.

\begin{figure}[htbp]
\begin{center}
\includegraphics[width=7cm,height=7cm,keepaspectratio,clip]{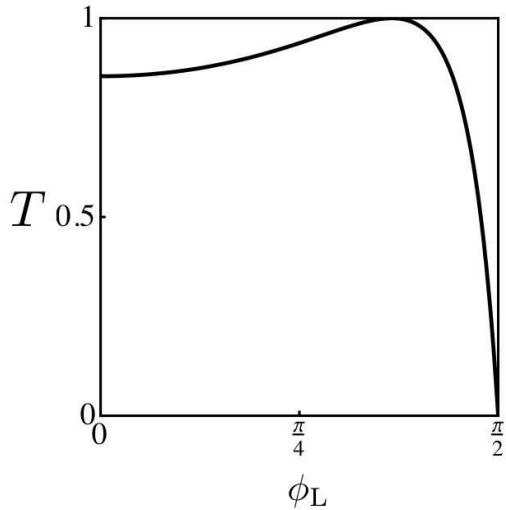}
\end{center}
\caption{ 
The transmission coefficient in the low energy limit as a function of the incident angle. 
We assume $V_{\rm L} = 0$ and $V_{\rm R} = 0.8$, 
consistent with the potential energy given in Eq. (\ref{PotNum}). 
}
\label{TPhiL.fig}
\end{figure}

The transmission coefficient $T$ in the low energy limit 
as a function of the incident angle $\phi_{\rm L}$ is shown in Fig.~\ref{TPhiL.fig}. 
We use $V_{\rm L} = 0$ and $V_{\rm R} = 0.8$ as in Eq. (\ref{PotNum}). 
The incident angle where the perfect transmission occurs is seen in Fig.~\ref{TPhiL.fig}, as discussed above. 

Moreover, we also find that the Bogoliubov excitation 
experiences the total internal 
reflection when the incident angle satisfies the condition $\phi_{\rm L} \geq \phi_{\rm L,c}$, 
where the critical angle $\phi_{\rm L, c}$ is 
defined by 
\begin{eqnarray}
\sin{\phi_{\rm L, c}} \equiv {c_{\rm L}}/{c_{\rm R}}, 
\label{TIR}
\end{eqnarray}
where $c_{\rm L} < c_{\rm R}$, {\it i.e.}, $V_{\rm L} > V_{\rm R}$. 

\begin{figure}[htbp]
\begin{center}
\includegraphics[width=8cm,height=8cm,keepaspectratio,clip]{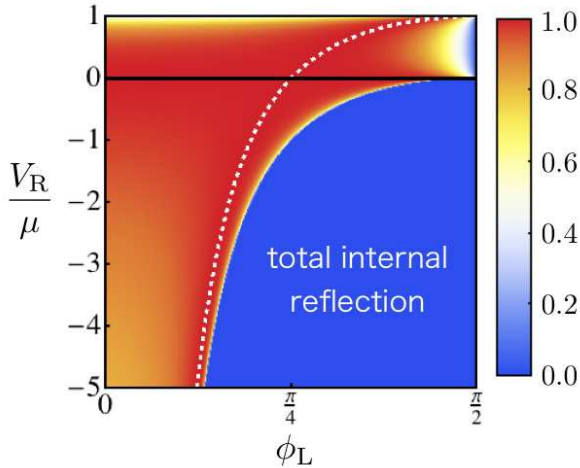}
\end{center}
\caption{
(Color online) 
The transmission coefficient in the low energy limit given by our analytical result 
is plotted in the $V_{\rm R}$ - $\phi_{\rm L}$ plane. 
We assume $V_{\rm L} = 0$. 
The solid line at $V_{\rm R} = 0$ represents the line where the ``ordinary'' anomalous tunneling occurs. 
The dotted line represents the line where the perfect transmission occurs 
in the same condition as the Brewster's law. 
There is a region where the total internal reflection occurs. 
}
\label{TPhiLVR.fig}
\end{figure}

In Fig.~\ref{TPhiLVR.fig}, 
the transmission coefficient $T$ in the low energy limit in Eq. (\ref{TRZ2}) 
is plotted in the $V_{\rm R}$ - $\phi_{\rm L}$ plane. 
We assume $V_{\rm L} = 0$. 
The solid line at $V_{\rm R} = 0$ represents the line where the ``ordinary'' anomalous tunneling occurs, 
{\it i.e.,} the perfect transmission occurs 
independently of the incident angle $\phi_{\rm L}$. 
On the other hand, the dotted line represents the line where the perfect transmission occurs 
in the same condition as the Brewster's law. 
As discussed above, 
there is a region where the total internal reflection occurs at $V_{\rm R} < V_{\rm L} = 0$, 
as seen in Fig.~\ref{TPhiLVR.fig}.

\subsection{Impedance Matching of Bogoliubov Phonon}

The Brewster's law and the total internal reflection of Bogoliubov excitations are naturally understood 
by recalling that expressions in Eq. (\ref{TRZ2}) for the transmission and reflection coefficients 
have the same forms as those of the electromagnetic wave~\cite{LandauElectromagnetic}. 
The transmission and reflection coefficients of an electromagnetic wave going 
from the medium $\rm L$ to $\rm R$ are given by
\begin{eqnarray}
\begin{array}{llllll}
%\lim\limits_{\varepsilon\rightarrow 0}
T &= & 
\begin{displaystyle}
\frac{4
Z_{\rm L}Z_{\rm R}
\cos{\phi_{\rm L}}\cos{\phi_{\rm R}}
}{
(
Z_{\rm R}\cos{\phi_{\rm L}}
+
Z_{\rm L}\cos{\phi_{\rm R}}
)^{2} 
}
\end{displaystyle}
\\
\\ 
%\lim\limits_{\varepsilon\rightarrow 0}
R 
&=  &
\begin{displaystyle}
\frac{
(
Z_{\rm R}\cos{\phi_{\rm L}}
-
Z_{\rm L}\cos{\phi_{\rm R}}
)^{2}
}
{
(
Z_{\rm R}\cos{\phi_{\rm L}}
+
Z_{\rm L}\cos{\phi_{\rm R}}
)^{2}
},  
\end{displaystyle}
\end{array}
\label{TRZ2Z}
\end{eqnarray} 
when the electric field of the incident electromagnetic wave is perpendicular to the incident plane. 
The symbols $Z_{\rm L}$ and $Z_{\rm R}$ denote the impedance of 
the electromagnetic wave 
$Z^{\rm el}=\sqrt{\epsilon'/\mu'}$ for each medium. 
Here $\epsilon'$ is the permittivity, and $\mu'$ is the magnetic permeability. 
When the magnetic permeability for two mediums are equal, 
$Z_{\rm R}/Z_{\rm L}$ becomes equal to $c_{\rm L}/c_{\rm R}$, 
where $c_{\rm L}$ and $c_{\rm R}$ are light speeds 
in left and right mediums. 
As a result, Eq. (\ref{TRZ2Z}) coincides with Eq. (\ref{TRZ2}). 
We note that the anomalous tunneling discussed in Refs.~\cite{Kagan2003,
Danshita2006,Kato2007,Kovrizhin2001} can be then simply regarded 
as an impedance matching between two identical condensates, 
where the impedance of the Bogoliubov phonon is inversely proportional to 
the speed of Bogoliubov phonon. 

It is worth mentioning that results in Eq. (\ref{TRZ2}) are different 
from the transmission and reflection coefficients of the classical sound wave 
at the interface between two mediums; 
at the interface, the transmission and reflection coefficients 
\begin{eqnarray}
\begin{array}{llllll}
%\lim\limits_{\varepsilon\rightarrow 0}
T &= & 
\begin{displaystyle}
\frac{4
Z^{\rm ac}_{\rm L}Z^{\rm ac}_{\rm R}
\cos{\phi_{\rm L}}\cos{\phi_{\rm R}}
}{
(
Z^{\rm ac}_{\rm R}\cos{\phi_{\rm L}}
+
Z^{\rm ac}_{\rm L}\cos{\phi_{\rm R}}
)^{2} 
}
\end{displaystyle}
\\
\\ 
%\lim\limits_{\varepsilon\rightarrow 0}
R 
&=  &
\begin{displaystyle}
\frac{
(
Z^{\rm ac}_{\rm R}\cos{\phi_{\rm L}}
-
Z^{\rm ac}_{\rm L}\cos{\phi_{\rm R}}
)^{2}
}
{
(
Z^{\rm ac}_{\rm R}\cos{\phi_{\rm L}}
+
Z^{\rm ac}_{\rm L}\cos{\phi_{\rm R}}
)^{2}
} 
\end{displaystyle}
\end{array}
\label{TRZac}
\end{eqnarray} 
of the classical sound are given by the same form as Eq. (\ref{TRZ2Z}). 
Here, $Z^{\rm ac}_{\rm L}$ and $Z^{\rm ac}_{\rm R}$ are 
the specific acoustic impedance $Z^{\rm ac}$ for each medium. 
$Z^{\rm ac}$ is defined as the ratio of the sound pressure to the particle velocity field 
and is given by $Z^{\rm ac}=\rho c$ with a mass density $\rho$ and a sound velocity $c$. 
The specific acoustic impedance for Bogoliubov excitations are given similarly 
from the calculation up to first order with respect to $u$ and $v$. 
The resultant specific acoustic impedance for Bogoliubov excitations is 
given by the product of the mass density of the condensate $(\propto 1-V_\nu)$ 
and the sound speed $(\propto |1-V_\nu|^{1/2})$  for each condensate $\nu=$L and R. 
Considering $c_\nu\propto |1-V_\nu|^{1/2}$ in Eq. (\ref{TRZ2}), 
it is obvious that Eqs. (\ref{TRZ2}) and (\ref{TRZac}) are different. 
%Thus, it can be understood that the transmission and reflection coefficients in Eq. (\ref{TRZ2}) in the present case 
%are described by another impedance. 

The difference of the impedance between the classical sound 
and the Bogoliubov phonon 
comes from the difference of the boundary condition; 
in the classical sound wave, which is the first sound, say, the hydrodynamic collective mode, 
the transmission coefficient is derived from continuous conditions 
at the interface 
for the velocity perpendicular to the interface and for the pressure~\cite{LandauFluid}. 
On the other hand, 
in the Bogoliubov phonon, which is the zeroth sound, 
say, the collisionless collective mode~\cite{1990NozieresPines, 1993Griffin, 2000Giorgini}, 
our result is derived from the constancy of the energy flux, 
and the property that the wave function of the Bogoliubov excitation in the low energy limit 
coincides with the macroscopic wave function of the condensate. 

\subsection{Comparison with Results on Tomonaga-Luttinger Liquids}

We shall discuss the relation between transmission and reflection coefficients 
obtained in our system and those obtained in one-dimensional boson systems. 
%Hohenberg showed that there does not exist long-range order in one and two dimensions 
%in Bose and Fermi systems for finite temperatures~\cite{Hohenberg1967}. 
%This result was derived by calculating the fluctuations of the order parameter 
%assuming a broken symmetry, and assuming infinite systems. 
%However, the trapped gases are of finite size, and hence 
%there is Bose-Einstein condensation for sufficiently low temperatures 
%in trapped Bose systems~\cite{Ho1998,Monien1998}. 
%As indicated in Ref.~\cite{Monien1998}, 
%considering the small fluctuation of the phase and density 
%around the saddle-point solution, 
%this system can be described by the Luttinger liquid Hamiltonian~\cite{Haldane1981PRL,Haldane1981JPC}. 
%The Luttiger liquid is an universality class, 
%describing the one-dimensional interacting boson systems and fermion systems 
%as the same model. 
%The difference between boson and fermion systems is contained in the Luttinger liquid parameter $K$. 
As mentioned in the introduction, 
the perfect transmission in the Tomonaga-Luttinger liquids was investigated 
extensively and intensively in the study of quantum wire. 
Using the renormalization-group, 
Kane and Fisher showed that 
the barrier is an irrelevant perturbation, say, the perfect transmission occurs, 
when the Luttinger liquid parameter $K$ satisfies $K > 1$~\cite{Kane1992}. 
In the fermion case, 
$K > 1$ corresponds to an attractive fermion system, 
$K = 1$ a free fermion system, 
and $1>K>0$ a repulsive fermion system~\cite{Giamarchi}. 
Safi and Schulz investigated the transport through a one-dimensional wire of interacting electrons 
connected to leads sufficiently longer than the wire~\cite{Safi1995}. 
%It was assumed that the Luttinger liquid parameter $K$ does not vary spatially in the leads. 
%Even if the Luttinger liquid parameter spatially varies in the wire, 
When one considers the multiple reflections, 
the transmission and reflection coefficients %$\tau$ and $\gamma$ 
are described by Luttinger liquid parameters in both leads, 
say, these coefficients do not depend on the Luttinger liquid parameter of the wire~\cite{Safi1995}. 
%regardless of %the Luttinger liquid parameter in 
%the wire~\cite{Safi1995}. 
Safi and Schulz concluded that the perfect transmission occurs 
when the Luttinger liquid parameters $K_{\rm L}$ and $K_{\rm R}$ of the left 
and right leads are the same 
{\it i.e.,} $K_{\rm L} = K_{\rm R} (>1)$. 
%where $K_{\rm L}$ and $K_{\rm R}$ are Luttinger liquid parameters in the left and right leads.   

Let us compare the transmission coefficient $\tau$ and 
the reflection coefficient $\gamma$ for the density fluctuation in the system treated in this paper 
with those in the one-dimensional system. 
These coefficients are defined by the ratio 
of density fluctuations of transmitted and reflected waves to the incident wave. 
In the Tomonaga-Luttinger liquids, 
the transmission and reflection coefficients $\tau$ and $\gamma$ with the multiple reflections, 
are given by $\tau=2U_{\rm L}K_{\rm R}/[U_{\rm R}(K_{\rm L}+K_{\rm R})]$, 
and $\gamma = (K_{\rm L}-K_{\rm R})/(K_{\rm L} + K_{\rm R})$. 
Here, $U_{\rm L}$ and $U_{\rm R}$ are speeds of collective excitations in the left lead and the right lead. 
%According to the universal class of the Luttinger liquids, 
%This result is also applied in the one-dimensional interacting boson systems. 
In the repulsive boson case, 
the Luttinger liquid parameter $K$ is larger than unity. 
$K =1$ corresponds to the Tonks-Girardeau gas~\cite{Tonks}. 
In the weakly interacting boson system with a short-range interaction, 
the Luttinger liquid parameter is given by $K_{\nu} = \pi\hbar \sqrt{\rho_{\nu}/mg}$, 
where $m$ is a mass, 
$g$ is a coupling constant of the short-range interaction, 
and $\rho_{\nu}$ is the ground state density with $\nu = {\rm L}$ for the left lead 
and with $\nu = {\rm R}$ for the right lead~\cite{Cazalilla2004}. 
The speed of the collective excitation $U_{\nu}$ in the weakly interacting boson system 
corresponds to that of the Bogoliubov phonon $c_{\nu}$, 
say, $U_{\nu} = c_{\nu} = \sqrt{g\rho_{\nu}/m}$~\cite{Lieb1963}. 
When the difference of Luttinger liquid parameters in both leads 
is caused by a potential step, 
the ground state density $\rho_{\nu}$ alone in both leads are different. 
As a result, the transmission and reflection coefficients $\tau$ and $\gamma$ with the multiple reflections 
are given by $\tau = 2c_{\rm L}/(c_{\rm L}+c_{\rm R})$ 
and $\gamma = (c_{\rm L}-c_{\rm R})/(c_{\rm L}+c_{\rm R})$. 
%$\rho_{0, {\rm L}}$ and $\rho_{0, {\rm R}}$ are ground density densities in the left and right leads. 
%$c_{\rm L}$ and $c_{\rm R}$ are sound speeds in the left and right leads. 
%The result given by Safi and Schultz~\cite{Safi1995} is obtained by the continuous condition 
%of the current and the momentum flux. 
%They defined the transmission coefficient as the ratio of the transmitted current to the incident current.  

%On the other hand, 
We shall consider the case for the normal incidence of the Bogoliubov phonon 
%
%the ratio $\tau_{\rm c}$ of amplitudes of the transmitted current 
%to the incident current 
up to first order with respect to $u$ and $v$, on the basis of our results in this paper. 
The transmission and reflection coefficients $\tau$ and $\gamma$ 
defined as ratios of amplitudes of the density fluctuation 
%of the transmitted and reflected waves to that of the incident wave. 
are given by $\tau = t $ and $\gamma = r$ 
in the low energy limit. 
%$\Psi_{\rm L}$ and $\Psi_{\rm R}$ are amplitudes of 
%the condensate wave function in the asymptotic regime, 
%and 
$t$ and $r$ are amplitude transmission and reflection coefficients in the low energy limit 
given in Eq. (\ref{TRZ}). 
As a result, we have expressions 
$\tau= 2c_{\rm L}/(c_{\rm L} + c_{\rm R})$, 
and $\gamma = (c_{\rm L}-c_{\rm R})/(c_{\rm L} + c_{\rm R})$, 
at normal incidence $\phi_{\rm L} = 0$. 
We note that this result agrees with the result given by the Tomonaga-Luttinger liquids as shown above.

In connection with the Tomonaga-Luttinger liquids, 
we shall comment on the difference between the reflection of the Bogoliubov excitation 
and the Andreev reflection. 
Dynamics of one-dimensional Bose liquids has been investigated in several papers, 
where the negative density reflection has been found to occur~\cite{Tokuno2007,Daley2007}. 
Tokuno {\it et al.} studied the dynamics 
of one-dimensional Bose liquids~\cite{Tokuno2007}.  
%and found the Andreev-{\it like} reflection 
%at Y-junctions~\cite{Tokuno2007}. 
In the Y-shaped potential, 
the incident packet splits into two transmitted packets 
running in two branches 
and one reflected packet running back in the branch. 
They found the negative density reflection. 
Daley {\it et. al.} studied the wave packet dynamics, 
in the one-dimensional optical lattice, 
propagating 
across a boundary in the interaction strength, 
%described by the Bose-Hubbard model assuming 
using the time-dependent density matrix renormalization group method~\cite{Daley2007}. 
They also discussed the negative density reflection. 
In these papers, the negative density reflection is called the Andreev-{\it like} reflection, 
in the sense that the negative density reflection is analogous to the reflection 
of hole-like excitations at the interface between super-normal conductors.

This Andreev-{\it like} reflection should be however 
distinguished from the Andreev reflection in the following sense.
If the condensed Bose system has 
the Andreev reflection, 
there should exist the retroreflection; when the Bogoliubov excitation runs toward the wall 
with an incident angle, the reflected excitation 
would go back the way one has come. 
In the weakly interacting Bose system separated by the potential wall, 
however, there exists no retroreflection as shown in Sec. II, 
and hence there exists no Andreev reflection. 
The reflection is just ordinary and obeys the law of reflection in Eq. (\ref{lawReflection}). 
This observation comes from the study of 
the tunneling problem between different densities 
and at oblique incidence 
in the weakly interacting three-dimensional Bose system. 
As a result, the Andreev-{\it like} reflection in Bose system 
in the one-dimensional Bose system 
is different from the Andreev reflection. 
%The result in Ref.~\cite{Tokuno2007} 
%is owing to the characteristic properties of the Y-junction 
%composed of the one-dimensional branches and 
%the Dirichlet boundary condition. 
%This result can be regarded as a kind of the fixed end reflection. 

\subsection{Future Problems}

The realization of the Bose-Einstein condensation in atomic gases~\cite{Anderson1995, Davis1995} 
has allowed us to investigate properties of Bogoliubov excitations experimentally. 
Using nondestructive phase-contrast imaging, 
it was found that the speed of sound induced by modifying the trap potential 
was consistent with the Bogoliubov theory~\cite{Andrews1997}. 
Using the Bragg scattering, 
the static structure factor of a condensed Bose gas was measured in the phonon regime~\cite{Stamper-Kurn1999}. 
It was also reported that the excitation spectrum of a condensed Bose gas agrees 
with the Bogoliubov spectrum using the Bragg scattering~\cite{Steinhauer2002}. 
On the other hand, 
the transmission and the reflection of Bogoliubov excitations 
are one of the issues that has not been investigated yet experimentally. 

It was discussed 
how to observe the anomalous tunneling in Ref.~\cite{Danshita2006}.
%
%Before closing this section, 
%we shall comment on 
%a proposal of the experiment how to observe the anomalous tunneling. 
%Details were discussed in Ref.~\cite{Danshita2006}. 
%The system in a 1D box trap was realized in Ref.~\cite{Meyrath2005}. 
%The potential barrier is produced by a blue-detuned laser, 
%and a potential step could also be realized 
%using a detuned laser beam shined over a razor edge~\cite{Seaman2005}. 
%Bogoliubov excitation with a sole wave number ${\bf k}$ could be produced 
%by making use of the Bragg pulse~\cite{Stamper-Kurn1999,Steinhauer2002}. 
%To make an incident excitation, 
%the condensed Bose gas in one side of the barrier 
%should be exposed to the Bragg pulse. 
%After making the excitation by Bragg pulse, 
%a hold time is needed till the Bogoliubov phonon experiences the transmission and reflection against the barrier. 
%After the hold time, transmission and reflection coefficients could be observed using the time-of-flight. 
%If the technique in Ref.~\cite{Meyrath2005} could be extended to 3D systems, 
%the reflection and refraction of the Bogoliubov phonon could also be observed. 
If one investigates the reflection and the refraction of Bogoliubov excitations, 
a potential step is needed. 
The potential step could be made 
by using a detuned laser beam shined over a razor edge~\cite{Seaman2005}. 
Bogoliubov excitation with a sole wave number ${\bf k}$ could be produced 
by making use of the Bragg pulse~\cite{Stamper-Kurn1999,Steinhauer2002}. 
An advantage of making use of the Bragg pulse 
to make Bogoliubov excitations 
is that one can produce an excitation with a sole wave number ${\bf k}$ 
having an arbitrary incident angle against the potential barrier. 

By making use of the local modification of the trap potential, 
an excitation can be also created. 
In this case, a density modification is composed of excitations with many modes, 
and hence the transmission and reflection coefficients should be estimated using a mode decomposition, 
in order to study the energy dependence of the transmission coefficient. 

If we use a box trap~\cite{Meyrath2005},  
these experiments are analogous to those of the reflection 
and the refraction of the wave at the interface between two different homogeneous mediums. 
When we use a harmonic trap, not a box trap, 
the problem is extended to a kind of the problem of the reflection and the refraction 
with a refractive index which changes spatially.

In the weakly interacting one-dimensional Bose system with a short-range interaction, 
it is known that the excitation spectrum agrees with that of the Bogoliubov excitation~\cite{Lieb1963}. 
In this one-dimensional system, 
we show that transmission and reflection coefficients also 
agree with those of Bogoliubov phonon in the long wave length limit, in the present paper. 
The result on the Tomonaga-Luttinger liquid suggest to us that the perfect transmission occurs 
in the symmetric system even if the interaction is strong. 
From this suggestion, 
not only in the weakly interacting three-dimensional Bose system 
but also in the strongly interacting three-dimensional Bose system 
beyond the mean-field treatment, 
the tunneling problem of the collective excitation should be then investigated 
theoretically and experimentally. 
The strongly interacting Bose gas can be made 
using the Feshbach resonance~\cite{Feshbach1958}. 
It is a problem whether the anomalous tunneling could be observed or not 
using experimental techniques mentioned above, 
in such a dilute ultracold gas under the control of the interaction strength.  
On the other hand, 
it is also a problem whether the anomalous tunneling could be observed 
or not in superfluid He-4 which is a high dense system and a strongly interacting Bose liquid.

Before closing this section, 
we shall propose a problem on the theoretical side. 
Recently, Tsuchiya and Ohashi 
studied tunneling properties of Bogoliubov phonons 
%in condensed Bose system 
taking notice of the quasi-particle current 
near the potential barrier~\cite{Tsuchiya2008}. 
They found that the quasi-particle current increases near the potential barrier 
inducing the supercurrent counterflow. 
%They proposed an interpretation that this increase explains the increase of the transmission coefficient of the Bogoliubov phonon. 
In order to conserve the total current, 
they use the Gross-Pitaevskii equation added in the anomalous average. 
This formulation brings the gapful excitation. 
Within their formulation with the gapful excitation, 
they did not confirm whether the anomalous tunneling occurs or not. 
On the other hand, 
it is reasoned that 
the fact that the wave function of the excited state in the low energy limit 
corresponds to the condensate wave function, 
say, the gapless excitation declared by the Hugenholtz-Pines theorem~\cite{Hugenholtz1959}, 
is necessary to the occurrence of the anomalous tunneling. 
On the basis of this idea, 
Kato {\it et al.} used the Popov approximation 
%which is useful to introduce the gapless excitation 
in order to show the occurrence of the anomalous tunneling even at finite temperatures~\cite{Kato2007}. 
However, 
this treatment is not also necessarily sufficient at finite temperatures, 
because the Popov approximation does not satisfy the conservation law. 
Hohenberg and Martin~\cite{Hohenberg1965} showed that 
the Ward-Takahashi relation~\cite{Ward1950,Takahashi1957} which guarantees the conservation law 
derives the Hugenholtz-Pines theorem. 
Within the theory satisfying the number conservation 
and the gapless excitation, for instance using formulations~\cite{Kita2006,Yukalov2006}, 
problems 
whether the anomalous tunneling occurs or not 
and how the quasi-particle current behaves 
still remain.

\section{Conclusion}
%We show that 
%the perfect tunneling occurs 
%when the sound velocities of Bogoliubov phonon 
%of two condensates separated by the potential wall 
%are equal. 
We find that 
the Bogoliubov phonon experiences the perfect transmission 
distinguished from the ``ordinary'' anomalous tunneling, 
when the incident angle satisfies the specific condition 
equal to the Brewster's angle, {\it i.e.}, 
the sum of the incident angle and the refracted angle is $\pi/2$. 
We also find that the Bogoliubov excitation experiences 
the total internal reflection. 
Introducing the impedance for the Bogoliubov phonon, 
the anomalous tunneling can be regarded as the impedance matching. 
In the weakly interacting Bose system with the short-range interaction, 
the transmission and reflection coefficients are consistent between 
the Bogoliubov theory and the Tomonaga-Luttinger liquid. 
The negative density reflection in the interacting condensed Bose system 
cannot be necessarily identified with the Andreev reflection.

\section{acknowledgment}
We thank H. Nishiwaki, K. Kamide, S. Tsuchiya, Y. Torii, I. Danshita, K. Tashiro 
and D. Takahashi for helpful discussions. 
This research was partially supported by the Ministry of Education, 
Science, Sports and Culture, Grant-in-Aid for 
Scientific Research on Priority Areas, 20029007.
This work is supported by a Grant-in-Aid 
for Scientific Research (C) 
No. 17540314 
from the Japan Society 
for the Promotion of Science. 
S. W. acknowledges support from 
the Fujyu-kai Foundation 
and the 21st Century COE Program at University of Tokyo.

\end{document}